\def\year{2024}\relax
\documentclass[letterpaper]{article} % DO NOT CHANGE THIS
\usepackage{aaai22}  % DO NOT CHANGE THIS
\usepackage{times}  % DO NOT CHANGE THIS
\usepackage{helvet}  % DO NOT CHANGE THIS
\usepackage{courier}  % DO NOT CHANGE THIS
\usepackage[hyphens]{url}  % DO NOT CHANGE THIS
\usepackage{graphicx} % DO NOT CHANGE THIS
\usepackage{natbib}  % DO NOT CHANGE THIS AND DO NOT ADD ANY OPTIONS TO IT
\usepackage{caption} % DO NOT CHANGE THIS AND DO NOT ADD ANY OPTIONS TO IT
\DeclareCaptionStyle{ruled}{labelfont=normalfont,labelsep=colon,strut=off} % DO NOT CHANGE THIS
\usepackage{algorithm}
\usepackage{algorithmic}
\usepackage{amsmath}
\usepackage{caption}
\usepackage{subcaption}
\usepackage{xcolor}

\usepackage{newfloat}
\usepackage{listings}
\floatstyle{ruled}
\newfloat{listing}{tb}{lst}{}
\floatname{listing}{Listing}

\usepackage[normalem]{ulem}
\useunder{\uline}{\ul}{}
\title{Analysis of Media Writing Style Bias through Text-Embedding Networks\thanks{}}
\author{
    Iain J. Cruickshank,\textsuperscript{\rm 1}
    Jessica Zhu,\textsuperscript{\rm 1, \rm 2}
    Nathaniel D. Bastian\textsuperscript{\rm 1}
}
\affiliations{
    \textsuperscript{\rm 1} Army Cyber Institute, United States Military Academy, West Point, NY\\
    \textsuperscript{\rm 2} University of Maryland, Baltimore, MD\\
    iain.cruickshank@westpoint.edu, jeszhu@umd.edu, nathaniel.bastian@westpoint.edu
}

\begin{document}

\maketitle

\begin{abstract}
With the rise of phenomena like `fake news' and the growth of heavily-biased media ecosystems, there has been increased attention on understanding and evaluating media bias. Of particular note in the evaluation of media bias is writing style bias, which includes lexical bias and framing bias. We propose a novel approach to evaluating writing style bias that utilizes natural language similarity estimation and a network-based representation of the shared content between articles to perform bias characterization. Our proposed method presents a new means of evaluating writing style bias that does not rely on human experts or knowledge of a media producer's publication procedures. The results of experimentation on real-world vaccine mandate data demonstrates the utility of the technique and how the standard bias labeling procedures of only having one bias label for a media producer is insufficient to truly characterize the bias of that media producer.
\end{abstract}

\section{Introduction}

The rise of fake news and heavily biased media ecosystems has contributed to many societal ills across the world. For example, biased media ecosystems have led to increased polarization in society and destructive phenomena like `truth decay' \cite{Kavanagh:2018}. As such, there is distinct importance to better understanding and evaluating media bias. When it comes to evaluating media bias, most citizens rely on third-party bias estimations, such as Adfontes or mediabiasfactheck.org. While these websites and the people behind them often do provide valid bias labels, they also require subjective interpretations of the media producers, as well as culturally specific evaluation procedures which may not translate to other media ecosystems (e.g., the methods used to evaluate bias in the United States may not translate to other nations, like India, for example), and can only scale at the rate that humans can evaluate media websites. As such, there has been increasing interest to find more computational approaches to evaluating media bias.

Of particular note in the evaluation of media bias is writing style bias, which includes lexical bias and framing bias. Various computational approaches exist for detecting writing style bias in media texts, with Natural Language Processing (NLP) techniques being the most common. However, most approaches are unsupervised due to the lack of labeled data sets and the difficulty in obtaining non-skewed labels. Despite this, recent studies have proposed supervised machine learning models for detecting informational bias, which has shown that additional context surrounding informationally-biased sentences aids in detection. Network-based approaches for analyzing writing style bias have also been proposed, including Centering Resonance Analysis and graph-theoretic approaches. However, most of these methods construct graphs at the individual text level rather than between different texts. It is also important to consider content redistribution in news production via news agencies, where media sources frequently share textual content. This phenomenon must be taken into account when comparing the writing style biases of different media sources.

In this paper, we propose a novel approach for detecting writing style bias in media texts that takes into account the shared content between different media sources. Our approach applies natural language similarity estimation and a network-based representation of the shared content between articles to perform bias characterization, and we evaluate its performance on a data set of news articles. We show that our proposed method can produce meaningful characterizations of writing style bias across any number of articles and media producers. We do so without having any knowledge of how those media domains produce their articles, which is frequently required for bias estimation by human subjective means. As part of our analysis, we also find that media domains' exhibited writing style biases can vary substantially based upon the event being reported on, which makes the standard bias labeling procedures of only having one bias label for a domain insufficient to truly characterize the bias of a media domain.

\section{Related Research}

Since the rise of phenomena like `fake news' and the growth of heavily-biased media ecosystems, there has been renewed attention on understanding media bias. Hamborg et al. recently highlighted the various ways by which media can be biased and the computational approaches that exist (or do not exist) for the different forms of media bias \cite{Hamborg:2019}. An important result of these recent bias studies is that there are often distinct elements of topical bias, or \textit{what} one chooses to talk about, as well as writing-style bias, or \textit{how} one chooses to talk about that topic, present in any given media space. Of particular note is the word choice --- or lexical --- and framing biases in the text; a text can be biased by both the words, or phrases, it chooses to use as well as the context of keywords in the text \cite{Hamborg:2019, DAlonzo:2022}. A recent sub-class of framing is \textit{Information Bias}, which is the conveyance of side information about the main event in the text in order to frame that main event in a certain way for the reader \cite{Fan:2019, Guo:2022}. As an example of the effect of framing, \citet{Kahneman:1984} performed a study whereby they presented the same information, but in two different ways (one with natural numbers and one with probabilistic estimates) and found distinct differences in respondents' perceptions of the information. As an example of writing style bias created through word choice, an author can substitute words with similar syntactic meanings, but distinctly different connotations when describing something \cite{DAlonzo:2022}. For example, one could say ``the government is ditching vaccine mandates" versus ``the government is removing vaccine mandates", where the words ``remove" and ``ditch" serve the same function in the sentences, but give very different connotations to the reader.

To date, most approaches to dealing with word choice and framing biases --- or writing style bias --- utilize NLP techniques, particularly in the form of unsupervised, text sentiment labeling \cite{Hamborg:2019, Cox:2021, Hamborg:2022, Hamborg:2021, Semeraro:2022}. Occasionally, this analysis is supplemented by topic modeling for event detection and clustering topics on the same article together \cite{Julinda:2014, Best:2005, Hamborg:2022, Jimenez:2022}. The preponderance of unsupervised approaches is largely due to two main reasons. First, there are few gold-standard, labeled writing style bias data sets \cite{Hamborg:2019, Shahid:2020}. Second, labeling bias is very difficult for humans to do, given inherent personal biases, and is also context-dependent (e.g., having a certain political bias will be relative to contexts like the country of the media source) \cite{Hamborg:2019, Shahid:2020, Sridharan:2022}. It is important to note, however, that there has been a number of recent works that have published data sets and proposed supervised machine learning models around the concept of informational bias \cite{Fan:2019, Van:2020, Chen:2020, Liu:2022, Guo:2022}. Of particular note is that these studies have generally found that the inclusion of additional context (surrounding sentence, topics of the article, etc.) around informationally-biased sentences within an article aids in the detection of those informationally-biased sentences. 

In addition to the preponderance of NLP-based approaches, there are also network-based approaches to writing style bias. Most notably, Centering Resonance Analysis constructs networks from a text by creating concept chunks, relating those chunks to each other by their nearness in the text, and then using network analysis techniques to analyze the bias \cite{Corman:2002}. Graph-theoretic approaches for prioritizing social media posts for fact-checking treat media bias (and other forms of misinformation) as a virus using epidemic spread modeling \cite{Smith:2022}. Finally, some recent work has proposed using sentence-to-sentence networks for bias prediction at the sentence level in media articles \cite{Guo:2022}. Yet, the proposed graph-centric approaches typically work within, constructing graphs of texts at the individual level, rather than between different texts \cite{Corman:2002, Semeraro:2022}. Overall, writing style bias remains an active area of research, with the overwhelming majority of techniques utilizing unsupervised NLP techniques, particularly combining sentiment labels with human analysis.

An important consideration in the analysis of writing style bias in media texts is the phenomenon of content redistribution in news production via news agencies. Media sources frequently share textual content between themselves; authorized copyediting is a fundamental component in the production of news \cite{Hamborg:2019, Boumans:2018}. Oftentimes this text reuse will take the form of less well-known news media organizations recycling content from major news agencies, like the Associated Press, and publishing that text \cite{Hamborg:2019, Boumans:2018, Kim:2009, Sanderson:1997}. This type of phenomenon has been investigated using techniques from Semantic Text Similarity analysis (e.g., plagiarism detection)  and is typically used to winnow down a collection of texts to just those that are unique (i.e., low-overlap in textual similarity to any other texts in the corpus) \cite{Hamborg:2019, Agirre:2016}. This phenomenon of text re-use, or indiscriminate text use, can also result in ``nut-picking" whereby certain words or phrases which have a distinct valence or sentiment in regards to a topic are used by media sources that generally have the opposite sentiment toward that topic (e.g., the use of the typically negatively connotated word of ``socialism" in a U.S. media article that actually supports a ``socialism") \cite{DAlonzo:2022}. The presence of such text use can challenge methods based on sentiment or valence to characterize writing-style bias. The common phenomenon of text reuse in media publication implies that when trying to compare the writing style biases of different media sources, one must do this analysis in the context of possibly large amounts of shared text between all of the media sources.  

\section{Methodology}

Given the difficulties with creating a labeled data set for analyzing writing style media bias \cite{Hamborg:2019, Shahid:2020}, we opted to approach this problem as an unsupervised machine learning problem. While it is possible to use proxy labels, like the domain bias label for the website from which a text emanates from \citet{Cruickshank:2021}, these are at best weak labels of the actual writing style bias present in the text. This latter point will become more evident in the analyses conducted further on in our paper. Additionally, the use of sentiment as a proxy for the writing style bias of a text presents its own issues that are not easily reconciled without additional human analysis \cite{Hamborg:2019, Hamborg:2022, DAlonzo:2022}. Finally, we present a method that is capable of working with data that potentially has large amounts of text reuse between news sources, as is common in media \cite{Boumans:2018}. 

\begin{figure*}[!ht]
    \centering
    \includegraphics[width=1.0\textwidth]{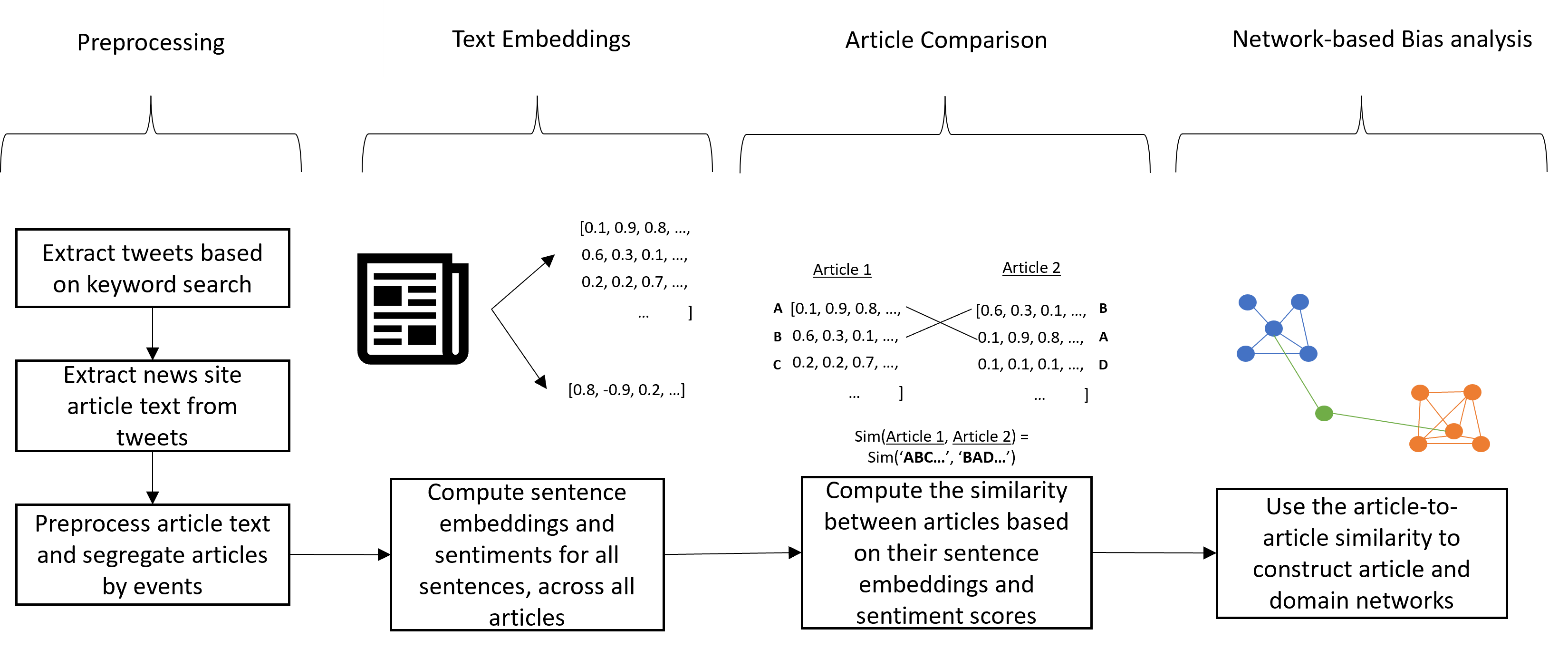}
    \caption{Proposed method for analyzing writing style media bias}
    \label{fig:method_diagram}
\end{figure*}

Thus, for the methodology in this paper, we use short-text embeddings and sentiment combined with carefully defined metrics over texts in order to construct networks for evaluating the writing style biases between texts and domains. Our proposed methodology for analyzing the writing style bias between articles and domains consists of four main steps: preprocessing and segregating the articles, creating sentence-level embeddings and the sentiment of each of the sentences for the articles, measuring article similarity by their sentence-level embeddings and sentiment differences, and then analyzing the articles and their domains by the networks created from the latter similarities \footnote{code and data available at: \url{ %https://anonymous.4open.science/r/analysis-of-media-writing-style-bias-through-text-embedding-networks-C7F5
https://github.com/ijcruic/analysis-of-media-writing-style-bias-through-text-embedding-networks
}}. Figure \ref{fig:method_diagram} displays our proposed methodology.

\subsection{Data Collection and Cleaning}

The first step in our proposed methodology is to collect and preprocess the textual data. We investigated the news surrounding military vaccine mandates in the United States. We chose to investigate this particular news story, as it was a major government policy with significant attention from partisan political sources, and thus a good topic for investigating media bias. We began by collecting the stories on military vaccine mandates that were shared over Twitter. By collecting news in this fashion, we can collect the articles that are most relevant to online discussions of the topic as well as acquire a greater variety of news sources than if we had collected only from a pre-determined list of news websites. We collected Twitter data from Twitter's search API\footnote{https://developer.twitter.com/en/docs/twitter-api} using the search phrase ``military vaccine mandate" from 1 February 2022 to 5 November 2022. In total, there were 1.3 million tweets of which 17.66\% had links to external websites within them. We then extracted the Uniform Resource Locators (URLs) from websites shared in the Tweets, deduplicated those URLs --- leaving 30,210 unique URLs --- and scraped the textual content of the URLs primarily using the Python package NewsPlease \cite{Hamborg2017}. This left 19,177 websites with textual content. From there, we labeled each successful article scraped from the URLs by their domain's political bias label \cite{Cruickshank:2021}, which left 7,065 articles with a known domain bias label.

Having obtained a selection of relevant news articles about a topic, we then preprocessed the articles. The first step of the preprocessing was to segregate the articles into distinct events. To do this, we adopted the topic modeling approach used in other works \cite{Julinda:2014, Best:2005}, which uses a technique like Latent Dirichlet Allocation \cite{Blei:2012} to segment the articles into distinct topics. It should be noted that other methods could be used, to include a Term Frequency Inverse Document Frequency (TF-IDF) and anchor approach used in \cite{Liu:2022}. We then further refined the segmentation of articles by down-selecting the articles within topics to those that only pertained to the primary event of the given topic (e.g., U.S. Air Force Academy refusing commissions to graduating cadets that refused the COVID vaccine, versus military members refusing the COVID vaccine). While this last step is not necessary to the overall methodology, we conducted it to ensure that the effects of other forms of bias (e.g., the bias of choosing what to report on \cite{Hamborg:2019}) are minimized relative to the writing style bias that we primarily wish to model and analyze. Table \ref{tab:dataset} summarizes the collected data set.

\begin{table}[!ht]
\begin{tabular}{|l|l|l|l|}
\hline
Primary Topic &
  \begin{tabular}[c]{@{}l@{}}Air Force \\ Cadets\\ Refused \\ Commision\end{tabular} &
  \begin{tabular}[c]{@{}l@{}}Navy \\ SEALs\\ Refuse \\ Vaccine\end{tabular} &
  \begin{tabular}[c]{@{}l@{}}Religious \\ Exemptions\\ to Vaccine \\ Mandate\end{tabular} \\ \hline
\begin{tabular}[c]{@{}l@{}}Number of\\ Articles\end{tabular} &
  45 &
  52 &
  72 \\ \hline
\begin{tabular}[c]{@{}l@{}}Number of \\ Unique\\ Domains\end{tabular} &
  41 &
  37 &
  51 \\ \hline
\end{tabular}
\caption{Summary of data set by events}
\label{tab:dataset}
\end{table}

Finally, we preprocessed the text within the articles by first removing junk sentences (e.g., sentences related to advertisement or subscription, ``clink the link to subscribe...''), breaking each of the texts into sentences, and adding the article title as a sentence. We broke the texts down to the sentence level because previous work has demonstrated the sentence as both being an important mesostructure of textual meaning and one that is often shared between news sources \cite{Kim:2009, Boumans:2018, Liu:2022}.

\subsection{Measuring Sentence Similarity}

Having obtained cleaned sentences for each of the articles, we then measure the similarity between the sentences of articles in order to compare those articles. We adopt a two-step procedure for this comparison: step one is to embed the sentences and compute their sentiment scores, and step two is to match sentences by their embeddings and sentiment scores. In the first step, we embed each of the sentences into a vector space using a pre-trained Bidirectional Encoder Representations from Transformers (BERT) model\footnote{https://huggingface.co/sentence-transformers/distiluse-base-multilingual-cased-v2}. Embedding at the sentence level, rather than the word or document level, allows for better identification of particular biases, like informational bias, as well as better means of comparing biases between documents \cite{Fan:2019, Van:2020, Liu:2022, Guo:2022}. We compute the sentiment score for each of the sentences using Valence Aware Dictionary and sEntiment Reasoner (VADER) \cite{Hutto:2014}. Differences in sentiment can help to distinguish between subtle word choice or punctuation manipulations for bias. See Appendix A for more details on the use of embeddings and sentiment and how they compare to other means of text comparison.

In the second step, with an embedding vector and sentiment score for each sentence, we compare sentences across articles for their semantic similarity and sentiment differences. In particular, we use embeddings to identify semantically similar sentences between articles, where sentences with only minor formatting or paraphrasing differences are mapped to nearby points in the same embedding space, but would not have been considered the same by exact text matching. Thus, the use of embeddings versus exact text matching allows for a more flexible model that can better handle real-world textual data. More specifically, we consider those sentences with a cosine similarity score of their embeddings above a certain threshold as matches. For two sentences, $s_1$ and $s_2$, they are a match if, 

\begin{equation}
    cos(emb(s_1), emb(s_2)) > \tau_1
\end{equation}

\noindent where $cos(.,.)$ is the cosine similarity between the embeddings of the sentences, $emb(s_1)$, and  $\tau_1$ is the threshold at which two sentences are semantically similar enough to be considered a match. For this study we set $\tau_1=0.7$.

We further compare differences in sentiment between semantically similar sentences to separate between those sentences which may be semantically close but have subtle differences in word choice or punctuation that give rise to a different connotation, and hence a different sentence. For two sentences, $s_1$ and $s_2$, they are considered sentimentally unrelated, and therefore not a match if, 

\begin{equation}
    |sent(s_1) - sent(s_2)| > \tau_2
\end{equation}

\noindent where $sent(.)$ is the sentiment score of a sentence and $\tau_2$ is how different the sentiment between two sentences can be before they are considered different sentences. For this study we set $\tau_2 =0.1$.

\subsection{Measuring Article Similarity}

Having obtained similarity measurements between the sentences of two articles, we can now compute the similarity between articles themselves. Since media articles often have high text reuse and the incorporation of reused text in an article can be done in subtle ways for writing style bias --- like changing the order of sentences or adding/omitting certain sentences to a reused text --- we measure the similarity between the articles in a two-step procedure. In the first step, we give a unique character to all of the unique sentences, by their embeddings. To determine uniqueness across articles we consider those sentences which have a similarity, $sim(.,.)$, greater than 0 to be matched, unique sentences. Each of these unique sentences is then assigned a unique character. For the unique characters, we use a large subset of around 1,000 characters of the UTF-8 set of characters, which has over a million possible unique characters, to ensure the alphabet can support sets of long articles with no character reuse. With the sentences mapped to characters, we represent each article as a string of those characters. In this way, each article's string preserves the order of the sentences in the article, as well as what sentences are shared between that article and the other article. We can then measure the similarity between two articles by the edit, or Levenshtein, distance \cite{Zhang:2017} between their string representations. We also note that our information-theoretic, embedding-based method of comparing articles also allows for an unordered comparison between the sentences, whereby one can use something like the Overlap Coefficient \cite{Vijaymeena:2016}, with the sentences that match between articles and those that do not match, to also compute article-to-article similarity.

\subsection{Creating Article-level and Domain-Level Networks}

Having obtained article-to-article similarities, we now construct article-level and domain-level networks to analyze the writing-style bias. With the article similarities obtained from the previous step, we construct a weighted network, where each node is an article and each link is weighted by the similarity between the endpoint articles. Due to how the similarity was measured between articles, this network can then be used to analyze writing-style bias, by clustering the network, as well as to understand key source articles, by using standard network analysis techniques (as has been done in other studies on semantic text similarity \cite{Hamborg:2019, Kim:2009}). 

From the article-to-article network, we can induce a domain-to-domain network by matrix algebra:

\begin{equation}
    D = A^{\circ \frac{1}{2}} S A^{\circ \frac{1}{2} T}
\end{equation}

\noindent where $D$ is a domain-to-domain network, $A$ is the domain-to-article bipartite network, and $S$ is the previously described article-to-article network. $\circ \frac{1}{2}$ is the Hadamard, or elementwise, square root function. This domain-to-domain network can be similarly analyzed for writing style bias at the domain level. 
    
\section{Results and Discussion}

In this section, we analyze the results of the writing style bias networks created on the collected data set. In the first part of the results, we analyze the networks produced by our proposed methodology, by analyzing their topology. In the second part, we analyze the clusters in the network relative to known domain biases \cite{Cruickshank:2021} using known cluster evaluation metrics (e.g., Adjusted Rand Index \cite{Hubert:1985}).

\subsection{Network Results}

\begin{figure*}[!ht]
     \centering
     \begin{subfigure}[b]{0.3\textwidth}
         \centering
         \includegraphics[width=\textwidth]{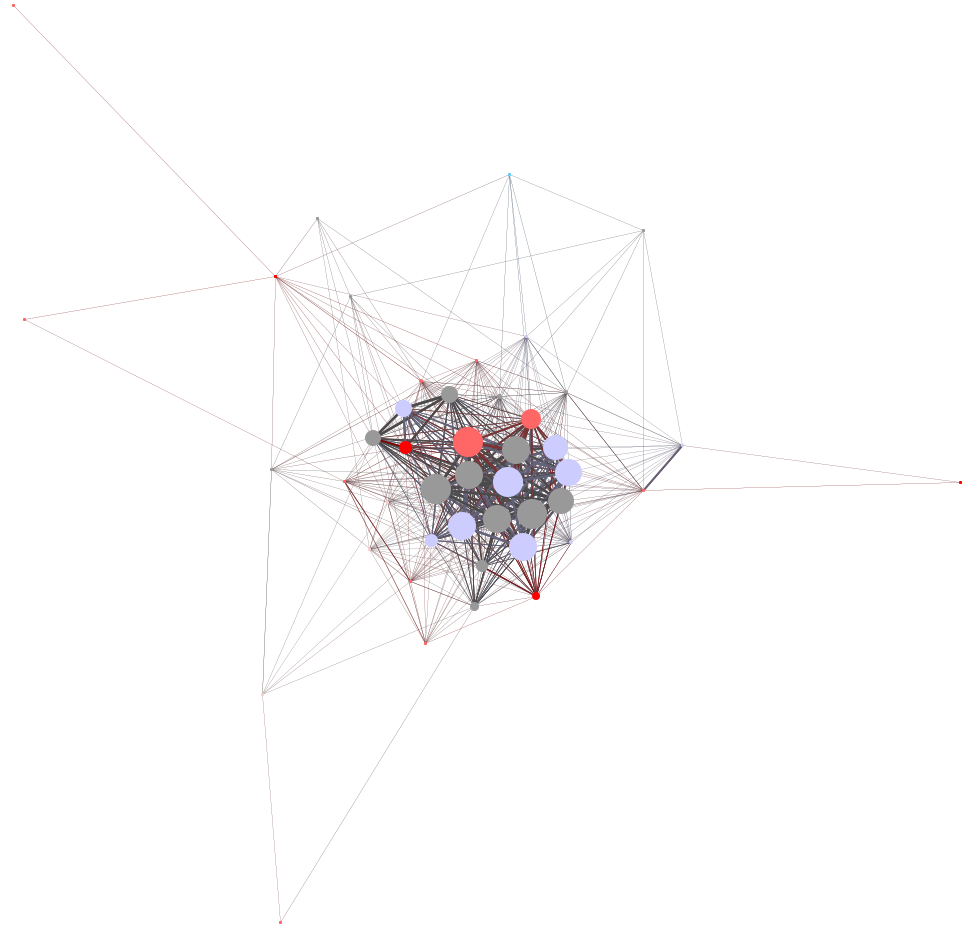}
         \caption{Air Force Cadets article similarity network}
         \label{fig:air_force_article_network}
     \end{subfigure}
     \hfill
     \begin{subfigure}[b]{0.3\textwidth}
         \centering
         \includegraphics[width=\textwidth]{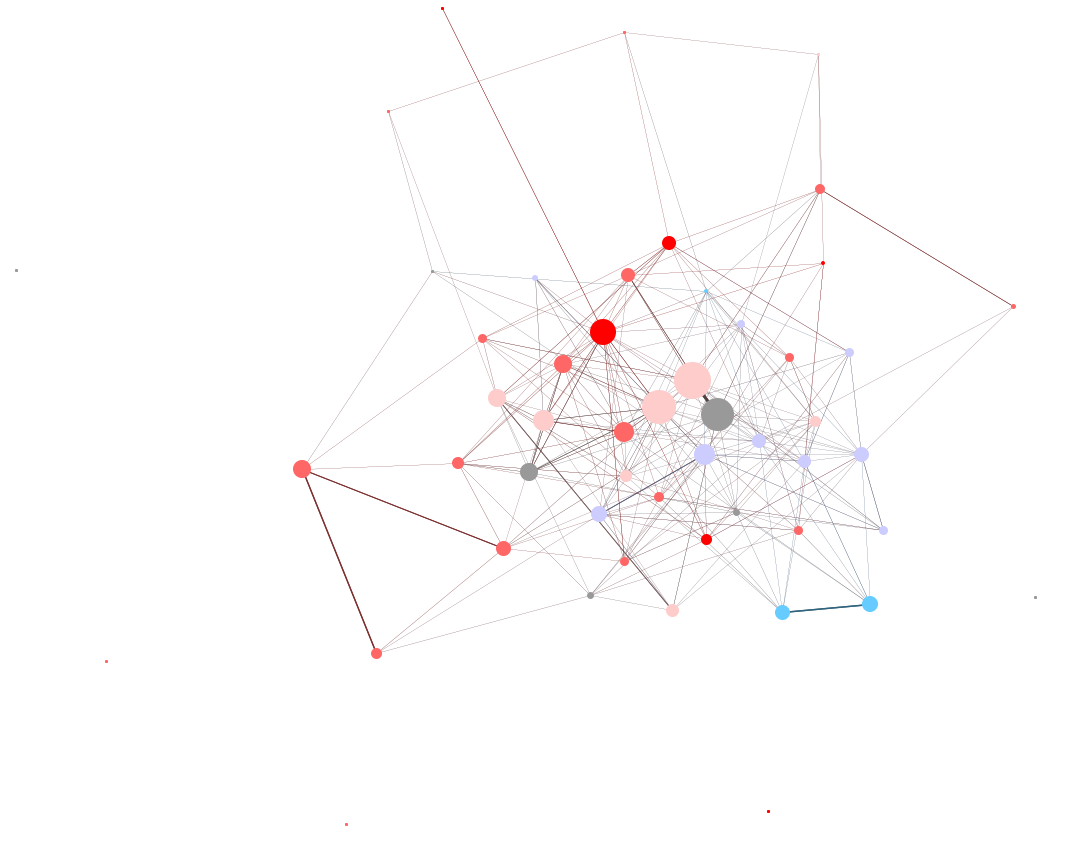}
         \caption{Navy SEALs article similarity network}
         \label{fig:navy_seal_article_network}
     \end{subfigure}
     \hfill
     \begin{subfigure}[b]{0.3\textwidth}
         \centering
         \includegraphics[width=\textwidth]{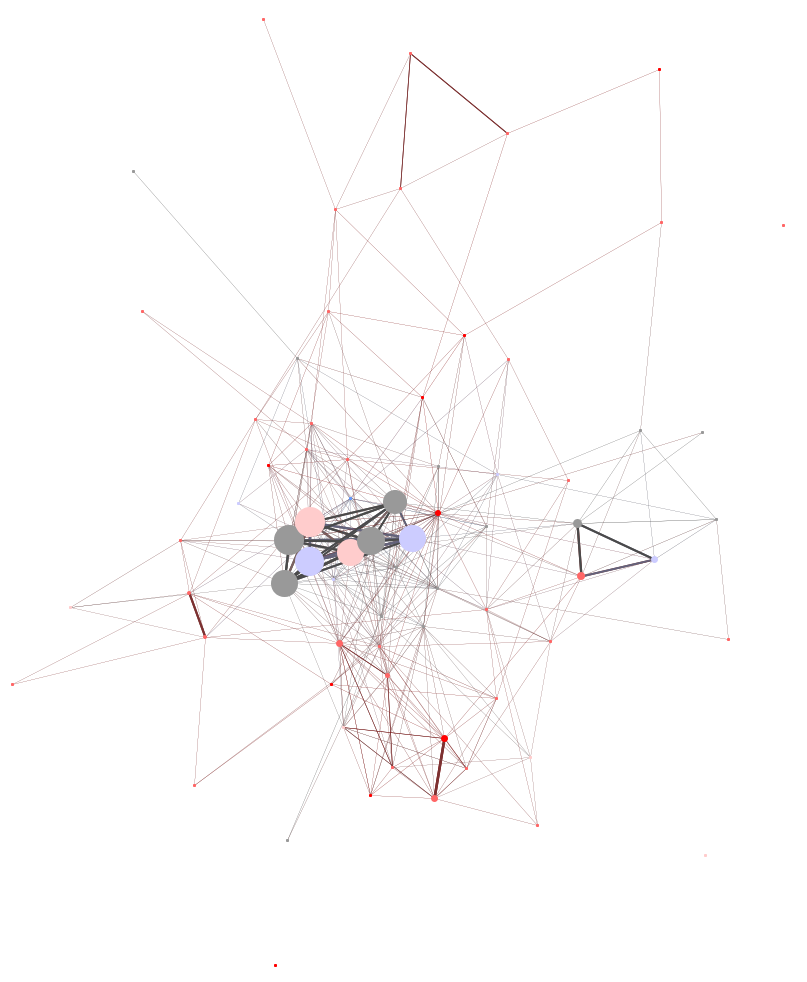}
         \caption{Religious Exemptions article similarity network}
         \label{fig:religious_exemptions_article_network}
     \end{subfigure}

     \begin{subfigure}[b]{0.3\textwidth}
         \centering
         \includegraphics[width=\textwidth]{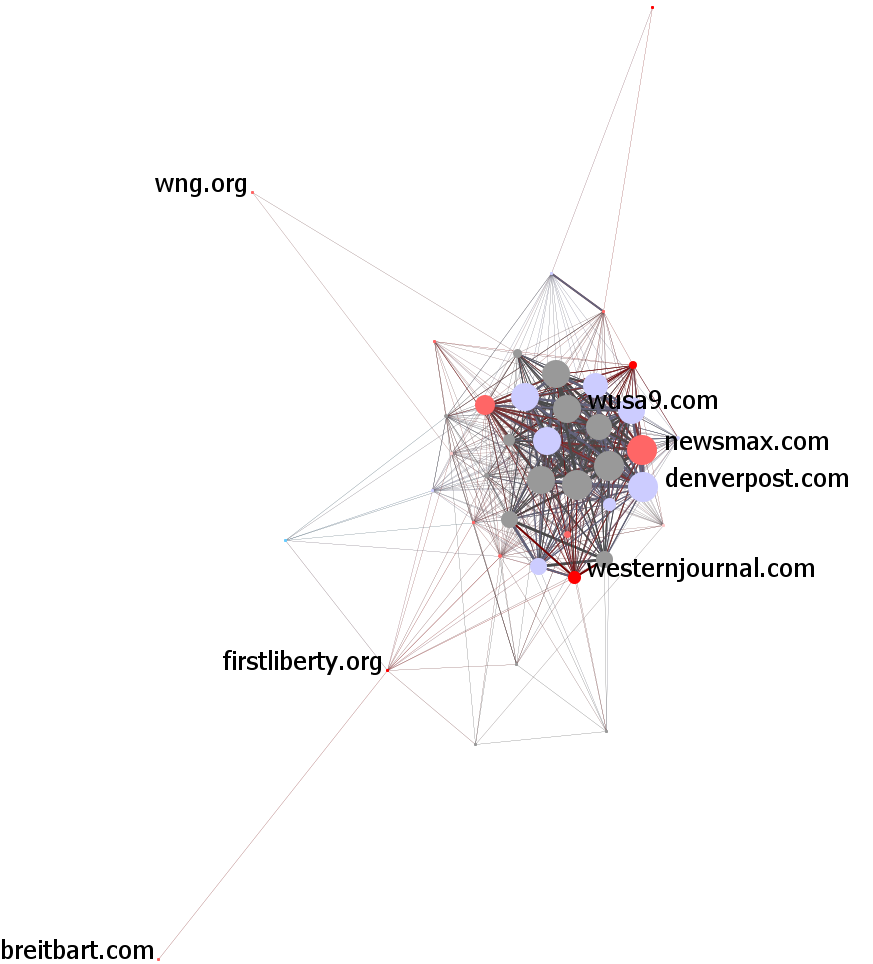}
         \caption{Air Force Cadets domain similarity network}
         \label{fig:air_force_domain_network}
     \end{subfigure}
     \hfill
     \begin{subfigure}[b]{0.3\textwidth}
         \centering
         \includegraphics[width=\textwidth]{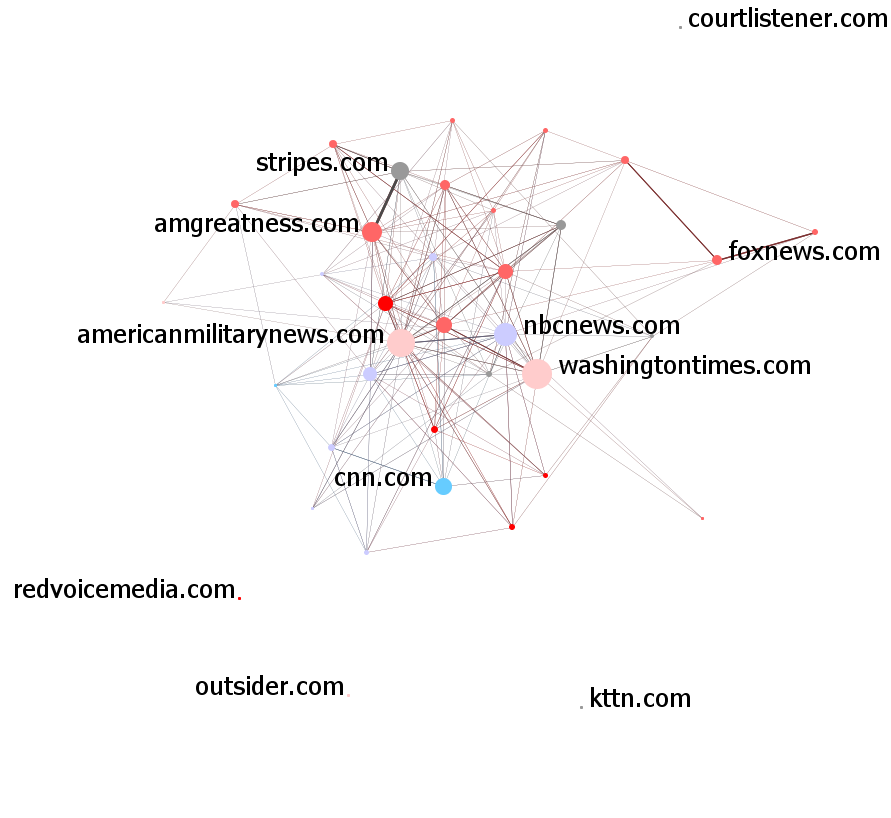}
         \caption{Navy SEALs domain similarity network}
         \label{fig:navy_seal_domain_network}
     \end{subfigure}
     \hfill
     \begin{subfigure}[b]{0.3\textwidth}
         \centering
         \includegraphics[width=\textwidth]{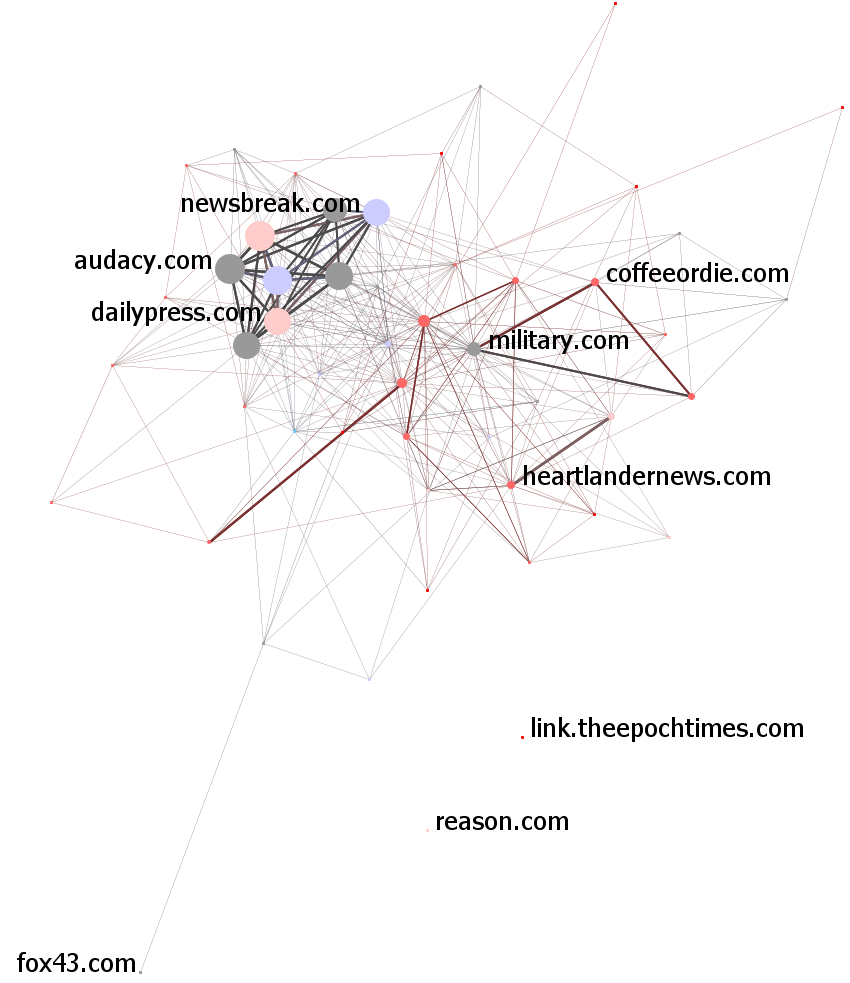}
         \caption{Religious Exemptions domain similarity network}
         \label{fig:religious_exemptions_domain_network}
     \end{subfigure}
     
    \caption{Article similarity networks, $S$, and domain similarity networks, $D$, for each of the events in the military vaccine mandates news data set. Nodes are sized by their degree centrality and colored by their domain bias ratings (\textcolor{red}{red}=right, \textcolor{gray}{gray}=center, \textcolor{cyan}{blue}=left). There are six bias ratings ranging from far right to far left. Edges are sized by their weight, which represents the similarity, between 0 and 1, between the nodes at the endpoints of the edge.}
    \label{fig:simialrity_networks}
\end{figure*}

After applying our method on the data set, we had six different networks: two networks for each event. For each event (e.g., Air Force Cadets being denied a commission), we compute an article-to-article similarity network, $S$, and a domain-to-domain similarity network $D$. The six networks are visualized in Figure \ref{fig:simialrity_networks}.

From observation of the similarity networks, a couple of key results emerge. First, we observe that the topologies between the article and domain networks are similar within events. This makes sense from the data as for any event, most domains only have a single article. So the articles often represent a proxy for the domains reporting the event. Second, the networks often have a core-periphery structure, where there are a few articles or domains, nodes, that have high similarity to one another (i.e., the core) and several  nodes that have some similarity to certain core articles (i.e., the periphery), but not to other nodes on the periphery. This structural pattern is especially true in the Air Force Cadets and Religious Exemptions events. As these networks are a representation of a media ecosystem, this result makes sense as there is often a high amount of text reuse in media ecosystems \cite{Boumans:2018}. Thus, the cores of these networks represent the main content describing the events, while the periphery contains less of that main content and, possibly, includes other related content (i.e., informational bias). Indeed, from inspection of the domains comprising the core structures and the most central nodes by degree centrality in each of the networks, many, but not all, of these domains are more mainstream news sources or agencies (e.g., ABC, Washington Times, NBC, AP, etc.).

From the networks, we also observe that there are differences in the media reporting on each of the events. For example, the Air Force Cadets event has a strong core-periphery structure, while the Navy SEALs event has a much weaker core network structure. The Religious Exemptions event has a strong core, but also a larger periphery structure with the semblance of smaller cores in the periphery. Upon manual inspection of the articles from each of these events, this pattern  matches the nature of the reporting on the events. For the Air Force Cadets event, most of the reporting centers around the official statements from the U.S. Air Force Academy on denying commissions, while the Navy SEALs articles use various elements of information, including personal interviews, mentions of other U.S. Navy vaccine-related court cases, in addition to official statements. In fact, both the Navy SEALs and Religious Exemption events contain both isolate articles and domains. These articles and their respective domains report on aspects of each event that no other source does. Some example sentences from the periphery of each of the data sets are given in Table \ref{tab:sentence_examples}. From this analysis, we demonstrate that the network structure captures differences in how the various media actors portray events and, by doing so, represents the framing and informational biases present in media reporting.

\begin{table*}[!ht]
\begin{tabular}{|l|l|l|}
\hline
Event & Core Sentence Example & Periphery Sentence Example \\ \hline
Air Force Cadets &
  \begin{tabular}[c]{@{}l@{}}Three cadets at the U.S. Air Force Academy \\ who have refused the COVID-19 vaccine will \\ not be commissioned as military officers but \\ will graduate with bachelor’s degrees, the \\ academy said Saturday.\end{tabular} &
  \begin{tabular}[c]{@{}l@{}}I have been very vocal against the United States \\ Air Force Academy (my alma mater) for its \\ corrupt teaching of cultural Marxism and its\\  Covid ‘vaccine’ mandates.\end{tabular} \\ \hline
Navy SEALs &
  \begin{tabular}[c]{@{}l@{}}Justices Clarence Thomas, Samuel Alito \\ and Neil Gorsuch dissented.\end{tabular} &
  \begin{tabular}[c]{@{}l@{}}The decorated Navy veteran appeared to recite \\ the speech from memory, emphasizing certain \\ phrases that are alarmingly relevant to the tyrannical \\ times we’re living in now.\end{tabular} \\ \hline
Religious Exemptions &
  \begin{tabular}[c]{@{}l@{}}The Air Force became the second military \\ service to approve religious exemptions to \\ the mandatory COVID-19 vaccine, granting \\ requests from nine airmen to avoid the shots, \\ officials said Tuesday\end{tabular} &
  \begin{tabular}[c]{@{}l@{}}Health experts said Novavax’s COVID-19 vaccine \\ could help address religious arguments made for \\ exemptions.\end{tabular} \\ \hline
\end{tabular}
\caption{Sentences characteristic of Core articles and Periphery Articles from the networks of each of the three events.}
\label{tab:sentence_examples}
\end{table*}

\subsection{Comparison of Network Clusters to Known Domain Biases}

From the networks, we also observe that there are often strong links between articles and domains with different known biases. To understand better the link between common bias labelings and the networks, we first clustered the networks using a common network clustering algorithm (i.e., Louvain \cite{Blondel:2008}). We then compared these clusters to publicly-available domain bias labels compiled from online sources like \url{mediabiasfactcheck.org} and \url{allsides.com}. There are six bias ratings ranging from far right to far left (i.e., far left, left, left-center, center, right-center, right, and far right). The following table, Table \ref{tab:bias_comparisons}, displays the results of comparing the found clusters to the domain bias labels.

\begin{table}[!ht]
\begin{tabular}{|l|l|l|l|}
\hline
 &
  \begin{tabular}[c]{@{}l@{}}Air Force \\ Cadets\end{tabular} &
  \begin{tabular}[c]{@{}l@{}}Navy \\ SEALs\end{tabular} &
  \begin{tabular}[c]{@{}l@{}}Religious \\ Exemptions\end{tabular} \\ \hline
\begin{tabular}[c]{@{}l@{}}ARI between \\ Clusters and \\ Bias Ratings at \\ the Article Level\end{tabular} & 0.062673 & 0.00213 & 0.02405 \\ \hline
\begin{tabular}[c]{@{}l@{}}ARI between \\ Clusters and \\ Bias Ratings at \\ the Domain Level\end{tabular}  & -0.00704 & 0.05441  & 0.12442 \\ \hline
\begin{tabular}[c]{@{}l@{}}Modularity of \\ Bias Ratings \\ at the Article \\ Level\end{tabular}            & -0.01091 & -0.0141 & 0.00186  \\ \hline
\begin{tabular}[c]{@{}l@{}}Modularity of \\ Bias Ratings \\ at the Domain \\ Level\end{tabular}             & -0.02142 & 0.08939  & 0.08742 \\ \hline
\end{tabular}
\caption{Comparison of network clusters to known Domain-level bias labels. The network clusters and Bias labels are compared by Adjusted Rand Index (ARI) \cite{Hubert:1985}. We also show the modularity of the bias labels on the networks, indicating how much domains or articles share content within their respective biases versus between biases.}
\label{tab:bias_comparisons}
\end{table}

From Table \ref{tab:bias_comparisons}, it can be clearly observed that the network clusters do not have much relation to the known domain bias labels. Many of the ARI scores are less than $0.1$ indicating little relation between these quantities. The only exception is the domain network for the Religious Exemptions event, which has an ARI of $0.124$. While this still is not an indicator of a strong relationship between the network clusters and the known domain biases, it does match the topology of the network. The Religious Exemptions domain similarity network has a core of mostly center-biased domains and then connectivity in the periphery between right-biased domains. We then computed the modularity of the known biases over the networks to get a measure of whether domains or articles tend to use content from similarly biased articles or domains or not. The low modularity scores for each of the networks indicate that domains and articles do not show a preference for using material from other articles or domains that share the same bias. Finally, we observe that there are differences between the events in terms of how much their clusters match known domain biases; the Religious Exemptions event tends to agree more with known domain biases, while the Air Force Cadets event bear little relation to them. This result may be due to how polarizing an event or topic is and how diverse the information coming from that event is. For example, much of the Air Force Cadet's event stems from official statements from the U.S. Air Force and was not as polarizing as an event like the court cases around the religious exemptions to the COVID-19 vaccine, as this event mentions the very divisive topic of religion.

We lastly analyzed the clusters across all of the events for their relation to known domain biases. To do so, we used cluster ensembling (i.e., Locally Weighted Bipartite Graph Partitioning Algorithm \cite{Huang:2017}) to cluster the clusters from each of the events across all of the domains present. For those domains that did not produce an article for a given event, and hence were not present in that event's clusters, we gave a new cluster label to them. The ARI between the ensemble clusters and known domain bias labels was $0.09155$. While still not a high ARI, it is more positive than most of the event ARIs, indicating that as we aggregate across sub-events, the clusters more closely resemble the known bias labels. This result mirrors that of other studies, which found there are nuances in traditional bias labels between topics, but that over a larger group of topics, these biases become more stable \cite{DAlonzo:2022}.

\subsection{Sensitivity Analysis}

We conducted an analysis of the two user-set parameters of the proposed method. More specifically, we wanted to see how much varying the semantic threshold (i.e. $\tau_1$) and the sentiment threshold (i.e. $\tau_2$) affected the results. To analyze the sensitivity of the method to these two parameters, we constructed article-to-article networks, $A$, across all three data sets while varying the values of $\tau_1$ and $tau_2$ between 0.1 and 0.99. We then measured the difference in the constructed networks by the normalized manhattan distance between the network adjacency matrices, which is given by the equation:

\begin{equation}
    d(A_i, A_j) =\frac{\sum{|A_i - A_j|}}{N(N-1)}
\end{equation}

\noindent where $N$ is the number of nodes, which makes $N(N-1)$ the total number of possible links in an undirected network. This measure indicates how different two networks are when the  parameters involved  (i.e., $\tau_1$ and $\tau_2$) in their construction are varied. Having computed the normalized Manhattan distances between all of the possible networks, we then averaged these across all of the data sets where $\tau_1$ and $\tau_2$ match and across all of the parameter values for each parameter's set of values (e.g., all of the $\tau_2$ values for each combination of $\tau_1$ values and vice versa). The following figure, Figure \ref{fig:sensitivity_analysis} displays the average distances between networks for each of the $\tau_1$ and $\tau_2$ combinations.

\begin{figure*}[!ht]
     \centering
     \begin{subfigure}[t]{0.4\textwidth}
         \centering
         \includegraphics[width=\textwidth]{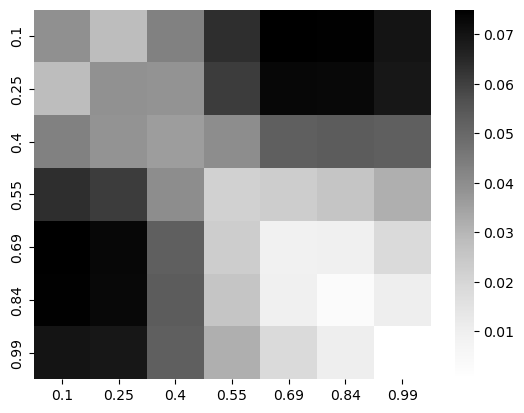}
         \caption{Sensitivity of $\tau_1$ (semantic sentence match threshold)}
         \label{fig:tau_1_sensitivity}
     \end{subfigure}
    ~
     \begin{subfigure}[t]{0.4\textwidth}
         \centering
         \includegraphics[width=\textwidth]{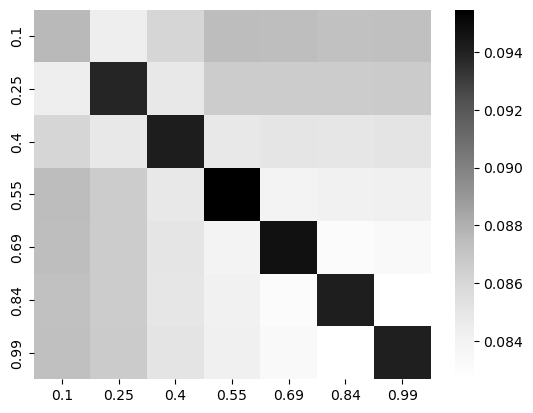}
         \caption{Sensitivity of $\tau_2$ (sentiment sentence match threshold)}
         \label{fig:tau_2_sensitivity}
     \end{subfigure}
     
    \caption{Normalized manhattan distances between networks produced using varying parameters. Each cell represents the average distance between networks constructed by the parameter value in the row and the parameter value in the column, averaged across all three data sets and all varying levels of the other parameter.}
    \label{fig:sensitivity_analysis}
\end{figure*}

From Figure \ref{fig:sensitivity_analysis}, it can be observed that the article-to-article networks produced are relatively stable for values of $\tau_1$ greater than 0.6 and for $\tau_2$ for values less than 0.3 or greater than 0.6. Thus, the method is relatively insensitive to reasonably high semantic thresholds ($\tau_1$), where the match between sentences should be at least 0.6 out of 1.0 for them to be considered a match and relatively insensitive to values of sentiment that are reasonably close to each other (e.g., not differing by more than 0.3). Based on these results, this validates our setting of $\tau_1=0.7$ and $\tau_2=0.1$, as these are within a stable range of results and represent having sentence pairs that are reasonably close semantically and only vary a little in sentiment to be considered as matches.

\section{Conclusion}

In this article, we proposed a new methodology for analyzing writing-style bias in media publications. Our methodology uses computational techniques to assess text overlap and identify bias, specifically at the sentence level, by applying natural language embeddings and a metric to evaluate text overlap between articles. The methodology does not rely on expert opinions or knowledge of media producer publishing practices but instead on what the domains produce. Our validation shows that text reuse and framing vary by the event, indicating that a single label for media producer bias is insufficient. Our study has some limitations, as we only focused on three events related to U.S. military COVID-19 vaccine mandates, and more research is needed to explore the generalizability of our findings across different topics and events. Additionally, we believe that there is an opportunity for future research to improve the methodology for determining whether two sentences are discussing the same thing in the same way.

\section*{Ethical Statement}
All articles collected for this study were done so under the provisions of Section 1078 of the U.S. Copyright Act and ensured that our collection action fell under the fair use category. The Tweets were collected in accordance with Twitter's terms of service at the time of collection.

\section*{Acknowledgements}

%This work was conducted within the Cognitive Security Research Lab at the Army Cyber Institute %at West Point and supported in part by the Office of Naval Research (ONR) under Support %Agreement No. USMA 20057. The views expressed in this paper are those of the authors and do not %reflect the official policy or position of the United States Military Academy, the United States %Army, the Department of Defense, or the United States Government.

% Use \bibliography{yourbibfile} instead or the References section will not appear in your paper
\bibliography{main}

\appendix
\section{Appendix A: Example of Text Manipulations and Similarity Measurements}

In this section, we demonstrate a number of approximate string comparison techniques, and how those techniques produce results around a number of ways sentences can be changed or manipulated, potentially for bias purposes. To demonstrate the techniques, we performed a number of manipulations on the base sentence of \textit{``Three cadets at the U.S. Air Force Academy who have refused the COVID-19 vaccine will not be commissioned as military officers but will graduate with bachelor\textbackslash's degrees, the academy said Saturday."}. The following Table \ref{tab:text_comparisons} summarizes the alterations to the base sentence.

\begin{table*}[!ht]
\begin{tabular}{|l|l|}
\hline
\multicolumn{1}{|c|}{Text Alteration} &
  \multicolumn{1}{c|}{New Sentence Text} \\ \hline
formatting typo &
  \begin{tabular}[c]{@{}l@{}}Three cadets at the U.S. Air Force Academy who have refused the COVID-19 \\ vaccine will not be commissioned as military officers, but will graduate with \\ bachelor\textbackslash{}'s degrees,\textbackslash n the academy said Saturday.\end{tabular} \\ \hline
inserted punctuation &
  \begin{tabular}[c]{@{}l@{}}Three cadets at the U.S. Air Force Academy who have refused the COVID-19 \\ vaccine will not be commissioned as military officers, but will graduate with \\ bachelor\textbackslash{}'s degrees, the academy said Saturday.\end{tabular} \\ \hline
missing unimportant word &
  \begin{tabular}[c]{@{}l@{}}Cadets at the U.S. Air Force Academy who have refused the COVID-19 vaccine \\ will not be commissioned as military officers, but will graduate with bachelor\textbackslash{}'s \\ degrees, the academy said Saturday.\end{tabular} \\ \hline
framing with quotes &
  \begin{tabular}[c]{@{}l@{}}Three cadets at the U.S. Air Force Academy who have \textbackslash{}"refused\textbackslash{}" the COVID-19 \\ vaccine will not be commissioned as military officers but will graduate with \\ bachelor\textbackslash{}'s degrees, the academy said Saturday.\end{tabular} \\ \hline
rephrasing &
  \begin{tabular}[c]{@{}l@{}}The Air Force Academy said Saturday that the three cadets who refused the \\ COVID-19 vaccine will not be commissioned as military officers but will \\ graduate with bachelor\textbackslash{}'s degrees.\end{tabular} \\ \hline
heavy rephrasing &
  \begin{tabular}[c]{@{}l@{}}The Air Force Academy is requiring cadets to vaccinate against COVID-19 \\ to commission.\end{tabular} \\ \hline
simple word change &
  \begin{tabular}[c]{@{}l@{}}Three cadets at the U.S. Air Force Academy who declined the COVID-19 \\ vaccine will not be commissioned as military officers, but will graduate with \\ bachelor\textbackslash{}'s degrees,\textbackslash the academy said Saturday.\end{tabular} \\ \hline
complex word change &
  \begin{tabular}[c]{@{}l@{}}Three cadets at the U.S. Air Force Academy who have refused the COVID-19 \\ vaccine will be denied commissions as military officers but will graduate with \\ bachelor\textbackslash{}'s degrees, the academy said Saturday.\end{tabular} \\ \hline
omission &
  \begin{tabular}[c]{@{}l@{}}Three cadets at the U.S. Air Force Academy who have refused the COVID-19 \\ vaccine will not be commissioned as military officers, the academy said \\ Saturday.\end{tabular} \\ \hline
unimportant addition &
  \begin{tabular}[c]{@{}l@{}}Three cadets at the U.S. Air Force Academy who have refused the COVID-19 \\ vaccine will not be commissioned as military officers but will graduate with \\ bachelor\textbackslash{}'s degrees, a spokesman from the academy said Saturday.\end{tabular} \\ \hline
unrelated &
  The U.S. Air Force Academy is located in Colorado Springs. \\ \hline
more similar unrelated &
  \begin{tabular}[c]{@{}l@{}}Three cadets from the U.S. Air Force Academy, along with other \\ commissioned military officers, presented at the graduate \\ academy colloquium on COVID-19 vaccination among bachelors this Saturday.\end{tabular} \\ \hline
foreign language &
  \begin{tabular}[c]{@{}l@{}}Tres cadetes de la Academia de la Fuerza Aérea de Estados Unidos que rechazaron\\  la vacuna COVID-19 no serán comisionados como oficiales militares, pero \\ se graduarán con una licenciatura, dijo la academia el sábado.\end{tabular} \\ \hline
\end{tabular}
\caption{Different types of text alterations which can happen both in a copy-editing scenario and when trying to inject bias into a sentence.}
\label{tab:text_comparisons}
\end{table*}

We then analyzed the many means of comparing texts using the base sentence and the altered sentences from Table \ref{tab:text_comparisons}. Specifically, we looked at exact text matching, approximate text matching by both the Jaccard Index and by Levenshtein distance, by the cosine similarity between whole-sentence embeddings and between pre-trained word embeddings, and by differences in VADER sentiment. For the approximate text matching by Jaccard Index, approximate text matching by Jaccard distance, and the cosine similarity between pre-trained word embeddings, we preprocessed the text following common convention with these techniques by removing common English stopwords, punctuation, and pronouns. The results of these comparisons are displayed in Table \ref{tab:text_comparison_results}.

\begin{table*}[!ht]
\begin{tabular}{|l|l|l|l|l|l|l|}
\hline
Text Alteration Technique &
  Exact Match &
  \begin{tabular}[c]{@{}l@{}}Jaccard \\ Approximate \\ Match\end{tabular} &
  \begin{tabular}[c]{@{}l@{}}Levenshtein \\ Fuzzy\\ Approximate \\ Match\end{tabular} &
  \begin{tabular}[c]{@{}l@{}}Sentence\\ Embeddings\\ Approximate\\ Match\end{tabular} &
  \begin{tabular}[c]{@{}l@{}}Cosine\\ Similarity \\ Between \\ Embeddings\end{tabular} &
  \begin{tabular}[c]{@{}l@{}}Difference in \\ VADER \\ Sentiment\end{tabular} \\ \hline
formatting typo          & FALSE & 0.94 & 0.99 & 1    & 0.99 & 0    \\ \hline
inserted punctuation     & FALSE & 1    & 0.99 & 1    & 0.99 & 0    \\ \hline
missing unimportant word & FALSE & 1    & 0.99 & 1    & 0.95 & 0    \\ \hline
framing with quotes      & FALSE & 1    & 0.99 & 1    & 0.98 & 0.15 \\ \hline
rephrasing               & FALSE & 0.94 & 0.82 & 0.98 & 0.97 & 0    \\ \hline
heavy rephrasing         & FALSE & 0.33 & 0.58 & 0.87 & 0.66 & 0.15 \\ \hline
simple word change       & FALSE & 0.83 & 0.95 & 0.99 & 0.99 & 0.15 \\ \hline
complex word change      & FALSE & 0.94 & 0.96 & 0.99 & 0.98 & 0.22 \\ \hline
omission                 & FALSE & 0.81 & 0.88 & 0.96 & 0.94 & 0.14 \\ \hline
unimportant addition     & FALSE & 0.94 & 0.91 & 0.99 & 0.98 & 0    \\ \hline
unrelated                & FALSE & 0.21 & 0.59 & 0.71 & 0.36 & 0.15 \\ \hline
more similar unrelated   & FALSE & 0.63 & 0.61 & 0.95 & 0.83 & 0.15 \\ \hline
foreign language         & FALSE & 0.02 & 0.47 & 0.06 & 0.97 & 0.14 \\ \hline
\end{tabular}
\caption{Results of different text comparison techniques across different text alteration techniques.}
\label{tab:text_comparison_results}
\end{table*}

From this example, one can observe that exact text matching is insufficient to handle comparing sentences generally. In terms of the approximate matching techniques, no particular one is clearly superior to the others when it comes to still matching minorly edited or paraphrased sentences and not matching those sentences which have been manipulated in ways that affect the bias or connotation of the sentence. The only exception is the ability of deep learning models to handle foreign languages; it's possible for a sentence to be in a foreign language and still be detected as similar to the base sentence by suitable, multilingual deep neural network embedding models. We can also observe that differences in sentiment can capture subtle changes in punctuation and word choice, which can be done to alter the connotation and bias of a sentence. Thus, from these results, we propose that one must compare both the semantics, in a flexible way, and the sentiment in order to best capture sentence-level bias alterations.

\end{document}